\documentclass[10pt,aps,twocolumn,showpacs,floatfix]{revtex4}
\usepackage{graphicx}
\usepackage{amsmath}
\usepackage{bm}
\usepackage{color}
\newcommand{\g}{\ifmmode\text{g}\else{g}\fi}
\newcommand{\q}{\text{'}}

\def \ddd  {{\rm d}}

\newcommand{\myparallel}{{\:\!\!\mkern3mu\vphantom{\perp}
\vrule depth 0pt\mkern1.7mu\vrule depth 0pt\mkern3mu}}
\def \dd  {{\partial}}
\newcommand{\chix}{\chi_{\raisebox{-1pt}{\scriptsize $x$}}}

\begin{document}

\title{Enhancement mechanism of the electron \g-factor in 
quantum point contacts}
\author{Gr\'egoire Vionnet}
\affiliation{School of Physics, The University of New South Wales,
  Sydney, NSW 2052, Australia}
\affiliation{Institute of Theoretical Physics, \'Ecole Polytechnique 
F\'ed\'erale de Lausanne (EPFL), CH-1015 Lausanne, Switzerland}
\author{Oleg P. Sushkov }
\affiliation{School of Physics, The University of New South Wales,
  Sydney, NSW 2052, Australia}

\date{\today}
\begin{abstract}
The electron \g-factor measured in a quantum point contact
by source-drain bias spectroscopy is significantly larger
than its value in a two-dimensional electron gas.
This enhancement, established experimentally in numerous studies,
is an outstanding puzzle. In the present work we explain the
mechanism of this enhancement in a theory accounting for the 
electron-electron interactions. 
We show that the effect relies 
crucially on the non-equilibrium nature of the spectroscopy
at finite bias.
\end{abstract}

\maketitle
A quantum point contact (QPC) is a narrow quasi-1D constriction 
linking two 2D electron gas (2DEG) reservoirs. 
It is essentially the simplest mesoscopic system
which makes it interesting both for technological applications and
on a fundamental level.
Experimental studies of QPCs started with the discovery of the
quantisation of the conductance in steps of 
$G_0=2e^2/h$~\cite{Wees1988,Wharam1988}, 
which is a single-particle effect well
understood theoretically~\cite{Buttiker1990}.
Many-body interactions/correlations in QPCs were first 
undoubtedly identified in the ``0.7-anomaly'' of the 
conductance and the \g-factor enhancement~\cite{Thomas1996},
and a few years later in the zero bias anomaly (ZBA) of 
the conductance~\cite{Cronenwett2002}.
Since their discovery, these effects have been the subject of numerous 
experimental
studies, see e.g. Refs.~\cite{Koop2007,Chen2009,Rossler2011,
Micolich2011,Burke2012}.
In spite of  20 years of studies there is no consensus about  the
mechanism of  the 0.7 \& ZBA. 
We believe that they are both due to the enhanced inelastic 
electron-electron scattering on the top of the QPC potential 
barrier~\cite{Sloggett2008,Lunde2009,Bauer2013}.
However, there are alternative theoretical models of these effects
based on various assumptions, see e.g. 
Refs.~\cite{Chuan1998,Spivak2000,Matveev2004}.
In the present work we do not address the 0.7 \& ZBA, but consider the 
mechanism underlying the electron \g-factor enhancement. 
We show  that a simple saddle-point potential model
combined with local electronic interactions 
is sufficient to capture the relevant physics.
There are two previous theoretical
works related to this problem, Refs.~\cite{Bauer2013,Chuan1996}.
Ref.~\cite{Chuan1996} considers the usual Landau Fermi liquid
exchange interaction mechanism of the g-factor enhancement in an
infinitely long quantum wire. This mechanism can hardly be relevant for a QPC
since the length of the quasi-1D channel connecting the leads is much shorter 
than the spin relaxation length.
Ref.~\cite{Bauer2013} addresses a real QPC and points out a magnetic splitting
enhancement  effect. While this effect does exist, we will show below 
that it is exactly cancelled out in a source-drain bias spectroscopy
experiment and therefore does not explain the observed phenomenon.

We consider the  conduction band electrons in a semiconductor.
Due to the spin-orbit interaction in the valence band, the
value of the single electron \g-factor can be very 
different from its vacuum value. For example in GaAs,
$\g_0 = -0.44$~\cite{Madelung1996}. 
 This value can be measured in fast processes, say in ESR, where
$\g_0$ is not renormalised by electron-electron 
interactions~\cite{yafet1963,Barnes}.
On the other hand, the static electron \g-factor $\g^*$
measured for example via static Pauli magnetisation in an infinite system
is enhanced compared to $\g_0$ due to the exchange electron-electron 
interactions~\cite{Janak1969,Chuan1996}. Considering that the time of flight 
of an electron through a QPC is of the order of the picosecond, how can the 
\g-factor be renormalised in such a fast process? We show below that the 
observed enhancement is specific to the source-drain bias spectroscopy 
method to measure the \g-factor in QPCs.

\begin{figure}[ht]
\includegraphics{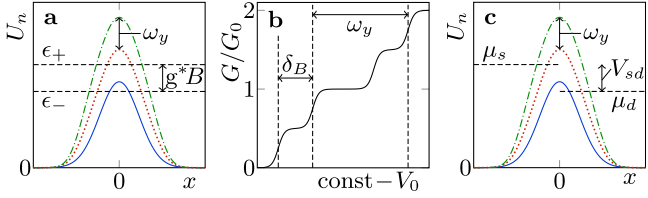}
\caption{(a)~Potential curves for the transverse 
channels $n=0$ (solid blue), $n=1$ (dotted red) and \mbox{$n=2$} 
(dash-dotted green),
and the magnetic field split chemical potential, 
$\epsilon_{\pm}$. (b)~Conductance in units of $G_0=2e^2/h$ versus the 
saddle point potential height $V_0$. (c)~Similar to~(a), but at zero
magnetic field and different chemical potentials in the source and
 drain reservoirs: $\mu_s-\mu_d=V_{sd}$.
}
\label{fig1}
\end{figure}

If we neglect the electron-electron interactions in the QPC the problem
can be described by the saddle point potential created by the gates
\begin{equation}
\label{spp}
V(x,y) = V_0 - \frac{1}{2}m\omega_x^2x^2 + \frac{1}{2}m\omega_y^2y^2 \ ,
\end{equation}
with $m$ the effective mass of the electron.
The electric current flows in the $x$-direction from the source to the 
drain.
The potential is separable and the QPC transmission problem is reduced to the 
solution of a one dimensional Schr\"odinger  equation with effective potential 
$U(x)$~\cite{Buttiker1990}. The potential is peaked at $x=0$ where
\begin{eqnarray}
\label{u0}
U_n(x)\approx U_{0n}-\frac{m\omega_x^2x^2}{2}\ ,  \ \
U_{0n}=V_0+\hbar\omega_y(n+1/2) .
\end{eqnarray}
Here $n=0,1,2...$ indicates the transverse channel.
At an infinitesimally small bias, the conductance is described by the 
transmission coefficient at the Fermi level.
Applying an in-plane magnetic field $B$ just spin splits the Fermi level
$\epsilon_{\pm}=\mu\pm \g^*B/2$ 
 with $\mu$ the chemical potential and the Bohr magneton set to unity.
The splitting is determined by the $\g^*$-factor 
which accounts for electron-electron exchange interactions in the leads.
The quasi-1D channel has no significant impact on $\g^*$
since its length ($\sim 100$nm) is much smaller than
the spin-relaxation length ($l_s\sim 10\mu$m). 
The energies $\epsilon_\pm$ and the potential curves describing the
QPC are sketched in Fig.\ref{fig1}a.
%%%%%%%%%%%%%%%%
%%%%%%%%%%%%%%%%%%%%%%%%%%%%%%%%%%%%%%%%%%%%%%%

\begin{figure}[ht]
\includegraphics{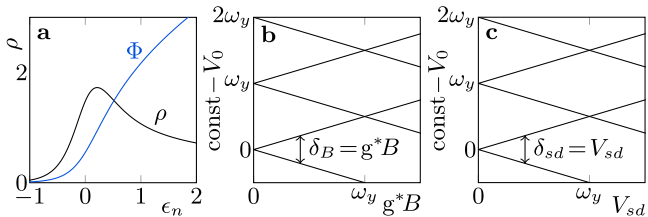}
\caption{(a)~Probability density at the top
of the barrier  $\rho$ and integral of the probability
$\Phi$ versus the electron energy. (b)~Position of the conductance steps
for non-interacting electrons in the QPC
 versus $B$ at infinitesimal $V_{sd}$. (c)~Similar to~(b) but 
versus the source-drain voltage $V_{sd}$ at $B=0$.
}
\label{fig2}
\end{figure}

An electron wave function  in a given transverse channel $n$
is a combination of incident, reflected and transmitted waves.
Near the peak of the potential, the wave function with energy $E_k=k^2/2m$ 
has the form (for $k \ge 0$)~\cite{Sloggett2008}
\begin{equation}
\begin{split}
\label{psi}
&\psi_{k,n}(x)\approx \left(\frac{mv_F^2}{2\omega_x}\right)^{1/4}
\varphi_{\epsilon_n}(\xi) \ ,
\\ 
&\varphi_{\epsilon_n}(\xi)=\sqrt{
\frac{e^{\pi\epsilon_n/2}}{\cosh(\pi\epsilon_n)}}
D_{\nu}(\sqrt{2}\xi e^{-i\pi/4}) \ , \\
&\epsilon_n=(E_k-U_{0n})/\omega_x \ ,  \qquad \xi=x\sqrt{m\omega_x} \ .
\end{split}
\end{equation}
Here $v_F$ is the Fermi velocity far from 
the barrier, 
$D_{\nu}$ is the parabolic cylinder  function, 
 $\nu=i\epsilon_n-\frac{1}{2}$ and $\hbar = 1$. 
The sign of $k$ indicates whether the electron is incident from
the left ($k \ge 0$) or from the right ($k \le 0$).
For our further analysis it is convenient to define the following
functions of energy
\begin{equation}
	\begin{split}
	\label{r}
&\rho(\epsilon_n)=|\varphi_{\epsilon_n}(0)|^2 \\
&\Phi(\epsilon_n)=\frac{\pi}{2\sqrt{2}}\int_{-\infty}^{\epsilon_n}
\rho(\epsilon^{\prime})d\epsilon^{\prime}
	\end{split}
\end{equation}
plotted in Fig.\ref{fig2}a.
%%%%%%%%%%%%
%%%%%%%%%%%%
Due to semiclassical slowing, the probability density at the top of the 
potential barrier, $|\psi_{k,n}(0)|^2\propto \rho(\epsilon_n)$,
is peaked at $\epsilon_n\approx 0.2$
strongly enhancing the interaction 
effects~\cite{Sloggett2008,Lunde2009,Bauer2013}.
The function $\Phi$ is proportional to the total electron density at $x=0$ 
in a given transverse channel.

Before considering the \g-factor enhancement, we illustrate the high 
susceptibility
of a QPC to a magnetic field $B$. Having the wave functions, it is easy 
to calculate the induced magnetisation $M(x,y)$
across the QPC which is directly relevant to the recent NMR 
experiment in Ref.~\cite{Kawamura2015}. Here we consider only the 
linear response to $B$. The only effect of electron-electron interactions
is to replace $g_0\to g^*$ (and renormalise $U_{0n}$ and $\omega_x$,
see below). The magnetisation in the leads is
$M(\infty,y)=\frac{m}{4\pi}\g^*B$ and the magnetisation
at the neck of the QPC is $M(0,0)=\frac{m}{4\pi}
\sqrt{\frac{2}{\pi}\frac{\omega_y}{\omega_x}}
\rho(\frac{\mu-U_{00}}{\omega_x})\g^*B$,
assuming that only the $n=0$ channel is open.
The latter  depends significantly on the energy through 
$\rho$ and the maximum enhancement is fairly large, 
$M(0,0)/M(\infty,0)\approx 2.5$ for $\omega_y/\omega_x \approx 3$. 
This single-particle effect is, however, unrelated to the \g-factor
measurement which we discuss in the next paragraph. 

The 
potential curves in Fig.\ref{fig1}a can be lowered and raised
by varying the QPC potential height $V_0$. 
 When the top of a potential curve crosses either one of
the $\epsilon_{\pm}$ horizontal lines, the conductance is changed by $G_0/2$. 
Each transverse channel leads to two split-steps separated by 
$\delta_B$ in $V_0$,
as illustrated in the plot of the conductance versus $V_0$ in Fig.\ref{fig1}b.
Without accounting for the electron-electron interactions in the QPC, the 
splitting is $\delta_B=\g^*B$.
 The standard way to represent Fig.\ref{fig1}b
is to plot the position of the steps versus magnetic field as shown
in Fig.\ref{fig2}b.
The slope of the lines in Fig.\ref{fig2}b is related to the g-factor 
and this is the basis for the \mbox{\g-factor} measurement. Of course, 
only the absolute value can be determined, $\g \to |\g|$.
Unfortunately, in experiments $V_0$ is unknown and
only the gate voltage $V_{g}$ is directly accessible.
$V_0$ is proportional to the gate voltage, $V_0=\alpha V_g$, and
a non-equilibrium method known as source-drain bias spectroscopy is 
used to exclude the unknown coefficient $\alpha$.
The magnetic field is set to be zero, but a finite bias $V_{sd}$
directly controlled experimentally
is applied across the QPC. The difference between the source and the drain 
chemical potentials is $\mu_s-\mu_d=V_{sd}$, as illustrated in
Fig.\ref{fig1}c. Similarly to the magnetic splitting case, when the 
top of a potential curve crosses  the $\mu_s$ or $\mu_d$ level, the 
differential conductance is changed by $G_0/2$.
Again, each transverse channel leads to two split-steps separated by 
$\delta_{sd}$ in $V_0$. For non-interacting electrons $\delta_{sd}=V_{sd}$.
The position of the steps versus  source-drain voltage is shown
in Fig.\ref{fig2}c.
The QPC \mbox{\g-factor} is
\begin{equation}
\label{gc}
\g_{Q}=\frac{(d\delta_B/dB)}{(d\delta_{sd}/dV_{sd})}=
\frac{(\partial V_g/\partial B)}{(\partial V_g/\partial V_{sd})} \ ,
 \end{equation}
where the derivatives are taken
at the same gate voltage. The unknown coefficient $\alpha$ is cancelled out
in the ratio in Eq.~(\ref{gc}).
Disregarding electron-electron interactions in the QPC,  $\g_Q=\g^*$.

Due to the many-body screening, the effective electron-electron Coulomb 
interaction is short-ranged and 
can be approximated by a $\delta$-function
\begin{equation}
\label{int}
V^c(x_1,x_2)=\pi^2 \lambda\sqrt{\frac{\omega_x}{m}}\delta(x_1-x_2) \ .
\end{equation}
Here we assume that the interaction is diagonal in transverse channels,
$\propto \delta_{n_1,n_2}$. In principle there is also an off-diagonal
interaction, but it 
does not influence our conclusions and is only relevant when several 
transverse channels are populated. The dimensionless coupling $\lambda$
is the four-leg vertex function which generally depends on $x$ and on the 
electron energy, $\lambda \to \lambda(x,\epsilon)$. For our purposes 
we need only $\lambda(x=0,\epsilon)=\lambda(\epsilon)$.
We have performed a random phase approximation (RPA) calculation of 
$\lambda(\epsilon)$ for the $n=0$ channel in GaAs
with $\omega_y/\omega_x=3$ and $\omega_x=1$meV, see Appendix~\ref{RPA}. 
It turns out that in the present case the results are
well approximated by discarding the energy-dependence and taking 
$\lambda=0.25$. 
This value agrees with the estimate $\lambda \sim 0.3/\sqrt{\omega_x}$ 
(with $\omega_x$ in meV) in Ref.~\cite{Sloggett2008}. We comment further 
on the $\epsilon$-dependence of $\lambda$ below.

\begin{figure}[ht]
\includegraphics{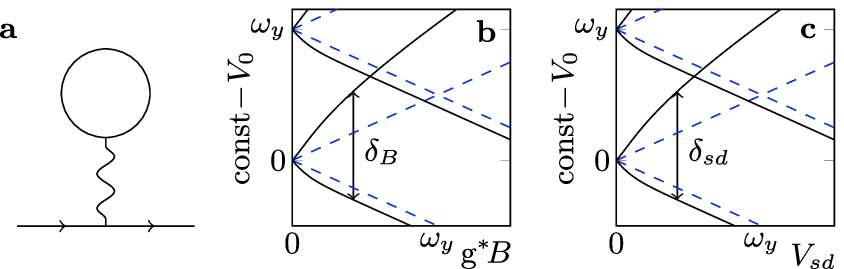}
\caption{(a)~Hartree self-energy 
diagram. (b)~Position of the conductance steps
 versus $B$ at infinitesimal $V_{sd}$ with the electron-electron interactions
 in the QPC accounted in the Hartree approximation.
 Black solid lines correspond to \mbox{$\lambda=0.25$} and 
 $\omega_y/\omega_x=3$.
Blue dashed lines correspond to the noninteracting case, 
$\lambda=0$. (c)~Similar to~(b) but
 versus the source-drain voltage $V_{sd}$ at $B=0$.
}
\label{fig3}
\end{figure}

The interaction leads to self-energy corrections to
the electron energy,
$
\epsilon_k \to \epsilon_k + \Sigma
$. We first consider the Hartree approximation, for which
the self-energy $\Sigma$ is given by the diagram shown in Fig.\ref{fig3}a.
This is equivalent to a self-consistent potential of electrons which gives
corrections to the height of the potential $U_{0n}$ and to $\omega_x$. We
focus on the former.
The potential at the top of the barrier 
is the sum of the potential $V_0$ created by the gates and the 
self-consistent potential created by the local density of electrons. 
In a magnetic field, the conductance steps arise when the top of a 
potential barrier in Fig.\ref{fig1}a touches a horizontal dashed 
line ('$\epsilon_+$' or '$\epsilon_-$').
Hence the conditions for the conductance steps are
\begin{equation}
	\begin{split}
	\label{tr1}
\q\epsilon_+\q: \ \: & c_n-\frac{V_0}{\omega_x}=-\frac{\g^*B}{2\omega_x}
+2\lambda\left[\Phi(0)+\Phi\left(-\frac{\g^*B}{\omega_x}\right)\right]\!
  \\
\q\epsilon_-\q:\ \: & c_n-\frac{V_0}{\omega_x}=\frac{\g^*B}{2\omega_x}
+2\lambda\left[\Phi(0)+\Phi\left(\frac{\g^*B}{\omega_x}\right)\right]\!
	\end{split}
\end{equation}
where $c_n$ is a constant that depends on the transverse channel.
The $\Phi(0)$-terms are due to interactions between electrons with
same spins, whereas the $\Phi(\pm \g^*B/\omega_x)$-terms are due to 
interactions
between electrons with opposite spins. The latter terms yield
an additional \mbox{$B$-dependence} compared to the non-interacting case. 
The position of the steps versus magnetic field which follow
from Eqs.\eqref{tr1} for $\lambda=0.25$ and $\omega_y/\omega_x=3$
are shown in Fig.\ref{fig3}b by black solid lines.
For comparison, the dashed blue lines show the non-interacting case,
identical to Fig.\ref{fig2}b.
%%%%%%%%%%%%%%%%%%%%%%%%%%%%%%%%%%%%%%%%%%%%%%%%%%%%%%%%%%
%%%%%%%%%%%%%%%%%%%%%%%%%%%%%%%%%%%%%%%%%%%%%%%
Fig.\ref{fig3}b indicates a very significant
Hartree enhancement of the splitting: $\delta_B > \g^* B$. 
In Ref.~\cite{Bauer2013} this effect was reported as enhancement
of the g-factor. However, let us look at how the source-drain normalisation
in Eq.~(\ref{gc}) influences the answer. The conditions for the 
source-drain conductance steps in the Hartree approximation
are
\begin{equation}
	\begin{split}
	\label{tr2}
\q\mu_s\q: \ \: & c_n-\frac{V_0}{\omega_x}=-\frac{V_{sd}}{2\omega_x}
+2\lambda\left[\Phi(0)+\Phi\left(-\frac{V_{sd}}{\omega_x}\right)\right]
  \\
\q\mu_d\q:\ \: & c_n-\frac{V_0}{\omega_x}=\frac{V_{sd}}{2\omega_x}
+2\lambda\left[\Phi(0)+\Phi\left(\frac{V_{sd}}{\omega_x}\right)\right] \ .
	\end{split}
\end{equation}
The factor $2$ in front of $\lambda$ in Eqs.(\ref{tr1}) and (\ref{tr2})
arises for different reasons. While in Eq.(\ref{tr1}) it is due to
the left-runners and the right-runners contributing to the density, 
in Eq.~(\ref{tr2}) it is due to the two spin polarisations. Due to 
the coincidence of the prefactors, Eqs.~(\ref{tr1}) and (\ref{tr2})
are identical upon the substitution $\g^*B \leftrightarrow V_{sd}$.
Hence, the plots of the position of the source-drain steps shown in 
Fig.\ref{fig3}c in 
black solid lines are identical to those in Fig.\ref{fig3}b.
(Again we show in blue dashed lines the non-interacting case).
Therefore in Eq.~(\ref{gc}) the ``enhancement'' is cancelled out
and $\g_Q=\g^*$.
There is no enhancement of the \g-factor measured
by source-drain bias spectroscopy due to the Hartree term.
Besides this analytical calculation, we have performed an
equilibrium self-consistent Hartree numerical calculation for a 
realistic QPC in a 3D geometry in the adiabatic approximation, 
see Appendix~\ref{scHartree}. 
This numerical calculation supports the above conclusion.

\begin{figure}[ht]
\includegraphics{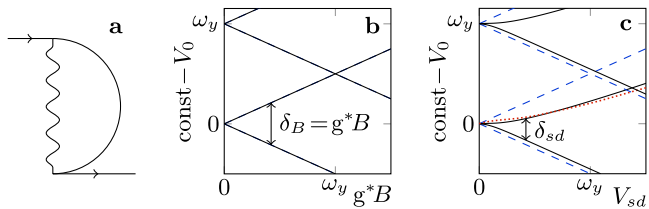}
\caption{(a)~Fock self-energy diagram. (b)~Position 
of the conductance steps versus $B$ at infinitesimal $V_{sd}$
with the electron-electron interactions
 in the QPC accounted in the Fock approximation.
Black solid lines correspond to $\lambda=0.25$ and $\omega_y/\omega_x=3$.
Blue dashed lines correspond to the noninteracting case, 
$\lambda=0$. (c)~Similar to~(b) but
 versus the source-drain voltage $V_{sd}$ at $B=0$.
The red dotted line accounts for the energy dependence
of $\lambda(\epsilon)$ in the RPA.
}
\label{fig4}
\end{figure}

We now account for the Fock exchange term, for which the 
self-energy diagram is plotted in Fig.\ref{fig4}a, and 
discard the Hartree self-energy.
Although in general this contribution to
the self-energy leads to a nonlocal potential,
in the $\delta$-function approximation (\ref{int}) it
becomes a local potential (generally spin-dependent).
We can therefore apply the same procedure as in the Hartree
case. However, 
the Fock self-energy is negative and for an electron with 
a given spin depends only on the density of electrons with the same spin.
Therefore,  the conditions for the conductance
steps in a magnetic field read
\begin{equation}
	\begin{split}
	\label{tr3}
\q\epsilon_+\q: \ \: & c_n-\frac{V_0}{\omega_x}=-\frac{\g^*B}{2\omega_x} 
-2\lambda\Phi(0)  \\
\q\epsilon_-\q:\ \: & c_n-\frac{V_0}{\omega_x}=
\frac{\g^*B}{2\omega_x}-2\lambda\Phi(0) \ .
	\end{split}
\end{equation}
It is very similar to the direct interaction case (\ref{tr1}), but
the sign of the $\lambda$-terms is opposite and there is
no term related to interactions between electrons with opposite spins.
Therefore the exchange contribution is \mbox{$B$-independent} and the 
position of 
the conductance steps, shown in Fig.\ref{fig4}b, are identical 
to the non-interacting  case, Fig.\ref{fig2}b. 
There is no exchange enhancement of the splitting, $\delta_B=\g^*B$.
%%%%%%%%%%%%%%%%%%%%%%%%%%%%%%%%%%%%%%%%%%%%%%%%%%%%%%%%%%
%%%%%%%%%%%%%%%%%%%%%%%%%%%%%%%%%%%%%%%%%%%%%%%
This is counterintuitive and very different from
what we know well about uniform systems~\cite{Janak1969}.
However, this does not imply that the g-factor measured by
source-drain bias spectroscopy is not changed. At zero magnetic field 
and finite bias, including the exchange contribution, 
the conditions for the conductance steps are
\begin{equation}
	\begin{split}
	\label{tr4}
\q\mu_s\q: \ \: & c_n-\frac{V_0}{\omega_x}=-\frac{V_{sd}}{2\omega_x}
-\lambda\left[\Phi(0)+\Phi\left(-\frac{V_{sd}}{\omega_x}\right)\right]
  \\
\q\mu_d\q:\ \: & c_n-\frac{V_0}{\omega_x}=\frac{V_{sd}}{2\omega_x}
-\lambda\left[\Phi(0)+\Phi\left(\frac{V_{sd}}{\omega_x}\right)\right] \ .
	\end{split}
\end{equation}
Compared to the Hartree case (\ref{tr2}),
the interaction contributions have opposite signs
and there is no factor 2 since only electrons with same spin
contribute. The position of the steps versus $V_{sd}$ is shown 
in Fig.\ref{fig4}c in black solid lines. For comparison, the red dotted line
shows a curve taking into account the energy dependence of $\lambda(\epsilon)$ 
obtained 
from the RPA calculation.  It is practically indistinguishable from 
the black solid line, thus justifying approximating $\lambda$ by a constant.
The exchange interaction reduces the splitting $\delta_{sd}$ compared to the
non-interacting case, $\delta_{sd}< V_{sd}$. 
Hence from Eq.~\eqref{gc}, the \g-factor is enhanced.
With parameters corresponding to the presented plots
($\lambda=0.25$, $\omega_y/\omega_x=3$) the g-factor enhancement is 
$\g_Q/\g^* \approx 1.5-2$ in agreement with experiments.
By increasing $\lambda$ one increases the enhancement, but then of course 
the single loop analysis becomes questionable.
When several transverse channels are populated, 
the screening must reduce the value of $\lambda$, thus reducing the
\g-factor enhancement in agreement with experiments~\cite{Thomas1996}.

We stress that $\g_Q$ is a multiplicative of $\g^*$ which is itself somewhat 
enhanced compared to $\g_0$ due to the exchange interaction in the leads. 
Implicit support for our analysis comes from the
experiments~\cite{Potok2002,Koop2008}. They measure the electron g-factor g
via magnetic field and temperature dependence of the magnetic field 
induced spin polarisation in QPC injection. This method does not rely on 
 source-drain bias spectroscopy and gives $\g \approx \g_0$. The fact
that $\g \neq \g_Q$ supports our analysis which is consistent with
$\g=\g^*\approx \g_0$, see also~\cite{com}.
 An explicit confirmation of our theory would come from a conductance 
measurement of the g-factor which
does not rely on the source-drain bias spectroscopy.

In conclusion, the g-factor enhancement in the Hartree-Fock 
approximation is
$\g_Q/\g^* = \frac{1+H}{1+H-F}\approx (1+F)$, where $H$ and $F$ are 
the Hartree and Fock electron-electron interaction contributions 
respectively. The numerator is due to the magnetic splitting 
and the denominator is due to the source-drain
normalisation.
Contrary to naive expectations, the exchange
Fock diagrams do not increase the magnetic splitting. However, the exchange 
diagrams  reduce the source-drain splitting in the non-equilibrium 
procedure used as normalisation of the energy scale, thus 
explaining the measured g-factor enhancement. The theoretical value 
of the enhancement is consistent with experiments.

We are grateful to A. P. Micolich, A. R. Hamilton, S. E. Barnes,
A. I. Milstein, and D. A. Ritchie
 for very important comments and discussions.

\onecolumngrid
\appendix
\pagebreak

\section{Self-consistent Hartree simulation in a realistic 3D geometry}
\label{scHartree}
\begin{figure}[ht]
 \includegraphics{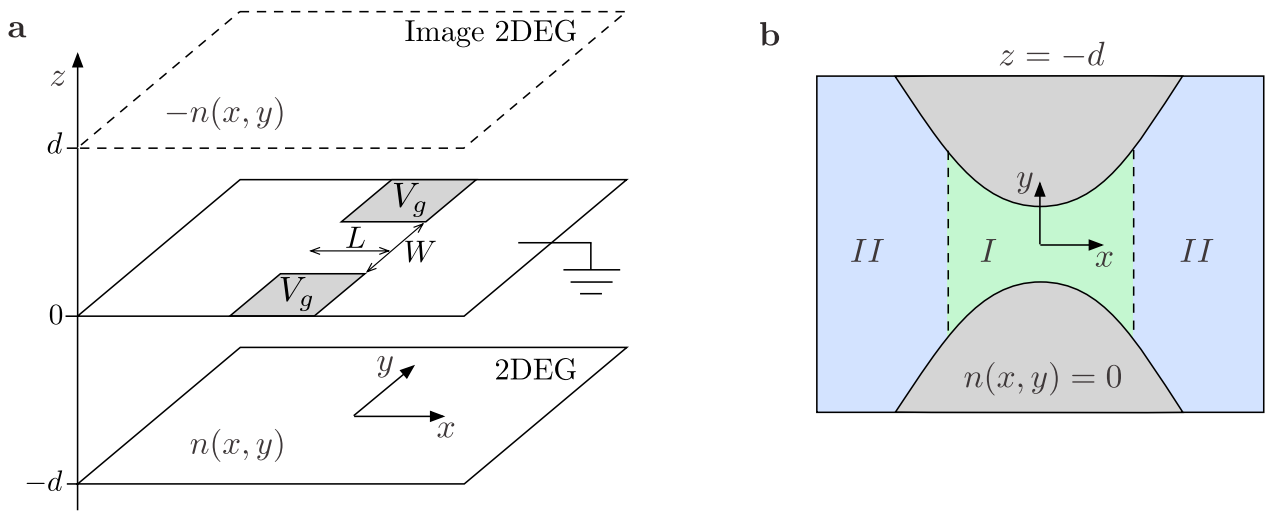}
\caption{
(a) Model of a QPC  as three equidistant two-dimensional layers: a 2DEG,
 a metallic plate with gates and an image 2DEG. 
At $z=0$, the gray metal gates have voltage $V_g$ and the remaining 
of this plane is grounded. The electron density in the  2DEG at 
$z=-d$ is $n(x,y)$, whereas in the image 2DEG at $z=d$ it is $-n(x,y)$. 
(b) Top view of the 2DEG plane ($z=-d$). The gray areas show the completely 
depleted regions where $n(x,y)=0$. The regions denoted $I$ (green) and 
$II$ (blue) are treated in an adiabatic and Thomas-Fermi approximation
respectively.
}
\label{Hartree3D}
\end{figure}
We consider here a simple electrostatic model of a QPC that takes into 
account the three-dimensional geometry of the experimental set-up. We 
focus on the Hartree interaction and ignore any exchange effect.
The system is approximated by three equidistant two-dimensional layers, 
as illustrated in Fig.\ref{Hartree3D}a.
The 2DEG is situated at $z=-d$, and its image is at $z=d$. The 
constriction is formed by applying a potential $V_g$ on the gray metal 
gates in the layer at $z=0$. We furthermore impose a zero potential 
outside of the gates in this plane. 
The total potential $U(x,y,z)$ is the sum of the electrostatic potential 
due to the metal gates and the Hartree potential $U_H(x,y,z)$ induced 
by the electrons  in the 2DEG and its image. In the plane $z=0$, the 
total potential reads
\begin{equation}
	\left. U\right|_{z=0} = U_{\rm gates} = \left\lbrace \begin{array}{ll}
	0 \qquad & \text{outside gates}\\
	-eV_g & \text{in gates}
	\end{array}\right. \ .
\end{equation}
Writing $\vec{r}_\myparallel=(x,y)$, the induced Hartree potential is
\begin{equation}
	U_H(\vec{r}_\myparallel,z) = \frac{e^2}{4\pi\varepsilon_0\varepsilon_r} 
	\int \ddd^2 \vec{r}_\myparallel\!{}' \ddd z' 
	\frac{ n(\vec{r}_\myparallel\!{}') }
	{\sqrt{|\vec{r}_\myparallel-\vec{r}_\myparallel\!{}
	\raisebox{0.5pt}{$'$}|^2  + (z-z')^2}} \left[ \delta(z'+d) 
	-  \delta(z'-d) \right] \ .
\end{equation}
This problem is simplest upon Fourier transforming the in-plane 
coordinates: $\hat{U}(\vec{q},z) = \int \ddd^2 \vec{r}_\myparallel 
U(\vec{r}_\myparallel,z)e^{i\vec{q}\cdot\vec{r}_\myparallel}$. A 
straightforward calculation yields
\begin{equation} \label{pot}
\left.	\hat U(\vec{q})\right|_{z=-d} = \hat U_{\rm gates} (\vec{q}) e^{-qd} 
+ \frac{e^2}{2\varepsilon_0\varepsilon_r}
\frac{\hat n(\vec{q})}{q}\left(1-e^{-2qd} \right)
\end{equation}
where $q=|\vec{q}|$.

We first describe the simulation at zero bias and and we then 
explain how we treat the non-equilibrium case.

\subsection{Zero bias: $V_{sd}=0$}
\begin{figure}[ht]
{
 \includegraphics{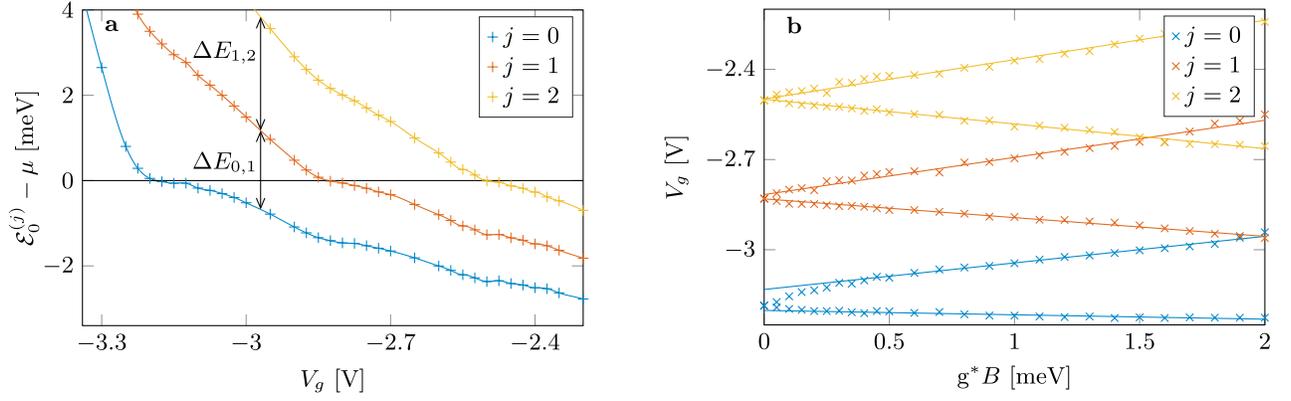}
}
\caption{
(a) Potential heights ${\cal E}^{(j)}_0$ of the transverse channels 
$j=0,1,2$ as a function of gate voltage $V_g$. The solid lines are guides 
to the eye. (b) Magnetic splitting for the conduction channels $j=0,1,2$ 
as a function of the Zeeman splitting  $\g^* B$. The solid lines are 
linear fits based on the data for $\g^* B > 0.4$ meV. 
}
\label{Hartree3Dres}
\end{figure}
We split the in-plane space $(x,y)$ in regions $I$ and $II$ as shown 
in Fig.\ref{Hartree3D}b and use different approximations for
the density $n(x,y)$ in each region. The region $I$ is defined
to be where the approximation $n_I(x,y)$ used in this region 
is valid. In region $II$, we use the Thomas-Fermi approximation
\begin{equation}
	n_{II}(x,y) = \frac{m}{2\pi}\sum_{\sigma=\pm 1} \Big(\mu + 
	\sigma \frac{\g^* B}{2} - U(x,y) \Big) \theta \Big(\mu + 
	\sigma \frac{\g^* B}{2} - U(x,y) \Big) 
\end{equation}
where $U(x,y)=U(x,y,-d)$ is the potential in the 2DEG plane.
In the region $I$, we use the adiabatic approximation and approximate 
the wavefunctions as $\Psi^{(j)}_k(x,y)=\psi^{(j)}_k(x)\chix^{(j)} (y)$ 
with $j$ labelling the transverse channels. The $\psi^{(j)}_k(x)$ are
1D scattering wavefunctions with the incident part having momentum $k$ 
asymptotically far from the constriction. Hence the sign of $k$ 
indicates the direction of the incident wave. 
We first solve 
numerically the one-dimensional Schr\"odinger equation at fixed $x$:
\begin{equation} 
	-\frac{1}{2m}\frac{\dd^2 \chix(y)}{\dd y^2} + U(x,y)\chix(y)  
	= {\cal E}_x\chix(y) \ .
\end{equation}
This yields for each $x$ a set of eigenfunctions $\chix^{(j)}(y)$ and 
eigenvalues ${\cal E}^{(j)}_x$ with $j=0,1,2\ldots$. The wavefunctions 
$\psi^{(j)}_k(x)$ are then found by solving 
\begin{equation}\label{schrox}
	-\frac{1}{2m}\frac{\dd^2 \psi^{(j)}_k(x)}{\dd x^2} + 
	{\cal E}^{(j)}_x \psi^{(j)}_k(x)  = \frac{ k^2}{2m} \psi^{(j)}_k(x)
\end{equation}
where the $x$-dependent transverse energies ${\cal E}^{(j)}_x$ 
play the role of an effective potential. 
Close to the constriction, we can approximate 
${\cal E}^{(j)}_x \approx {\cal E}^{(j)}_0 - 
\frac{1}{2}m\left(\omega_x^{(j)}\right)^2x^2 $
for which we know the exact solution as a function of the parabolic
cylinder functions. The validity of this 
harmonic approximation defines the region $I$. 
The density in that region is
\begin{equation} \label{dens}
\begin{split}
n_I(x,y) &= \sum_{\sigma=\pm 1} \sum_{j} \sum_{-\infty<k<\infty} 
|\psi_k^{(j)}(x)|^2|\chix^{(j)}(y)|^2 \theta\Big(\mu+\sigma 
\frac{\g^* B}{2}-\frac{k^2}{2m}\Big) \ .
\end{split}
\end{equation}
The densities $n_I$ and $n_{II}$ match pretty smoothly at the boundaries
between regions $I$ and $II$.

The self-consistent equation \eqref{pot} is solved iteratively with
the parameters (see Fig.\ref{Hartree3D}a): $L=300$nm, $W=500$nm, $d=90$nm,
$\mu=6$meV, $\epsilon_r=13$ and 
$m=0.07m_e$. This yields the density $n_\infty= 1.75 \cdot 10^{11}$ 
cm${}^{-2}$ asymptotically far from the constriction.
The potential heights ${\cal E}^{(j)}_0$ for the three first transverse 
channels are shown as a function of gate voltage in 
Fig.\ref{Hartree3Dres}a. The energy levels are roughly equidistant, 
$\Delta E_{0,1}(V_g) \approx \Delta E_{1,2}(V_g)$ 
(see Fig.\ref{Hartree3Dres}a), and 
evolve slowly with $V_g$, in agreement with experiments. 
The magnetic splitting in this self-consistent Hartree simulation is shown 
in Fig.\ref{Hartree3Dres}b. It shows a linear splitting, at least for 
large enough $B$, as observed experimentally. Comparison with the bare 
electrostatic potential from the gates, i.e. disregarding $U_H$, we find 
a largely enhanced magnetic splitting: $\delta_B \sim 10~\g^*B$.

\subsection{Finite bias: $V_{sd} \neq 0$} \label{sec:ooe}
\begin{figure}[t]
{
 \includegraphics{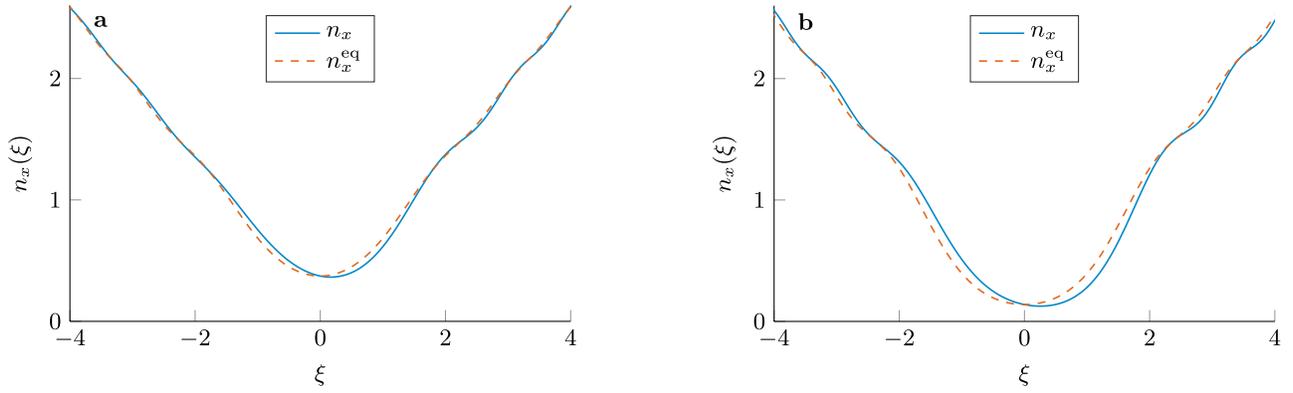}
}
\caption{
Comparison between the out-of-equilibrium 1D density $n_x$ and equilibrium
approximation $n_x^{\rm eq}$, using as unit length the
dimensionless $\xi=\sqrt{m\omega_x}x$. The chemical potentials are:
(a) $\mu_s-{\cal E}_0 = 0.4\omega_x$ and $\mu_d-{\cal E}_0 = 0$; 
(b) $\mu_s-{\cal E}_0 = 0$ and $\mu_d-{\cal E}_0 = -0.4\omega_x$.
}
\label{densOEQ}
\end{figure}
We restrict the discussion here to zero magnetic field. In the region $I$, we
have for $V_{sd}\neq 0$
\begin{equation}
	n_I(x,y) = 2 \sum_{j} |\chix^{(j)}(y)|^2\Big\lbrace \sum_{k > 0}
 |\psi_k^{(j)}(x)|^2 \theta(\mu_s - \frac{k^2}{2m})
 +  \sum_{k<0} |\psi_k^{(j)}(x)|^2 
 \theta(\mu_d - \frac{k^2}{2m})\Big\rbrace 
\end{equation}
where the factor $2$ comes from the spin degeneracy.
However, we cannot write such an out-of-equilibrium expression for the density
in the region $II$ since the Thomas-Fermi approximation assumes local
equilibrium. Therefore our method cannot 
treat a non-equilibrium situation and we need a simplifying approximation.
We are mostly
interested in what happens close to the top of the potential, i.e. for small
$|x|$. We therefore average symmetrically around $x=0$, 
\begin{equation}\label{subst}
	n(x,y) \to \frac{1}{2}(n(x,y)+n(-x,y)) \ .
\end{equation}
To justify this substitution, let's consider the one-dimensional density
for a single transverse channel
\begin{equation}
	n_x(x)=\sum_{k > 0}
 |\psi_k(x)|^2 \theta(\mu_s - \frac{k^2}{2m})
 +  \sum_{k<0} |\psi_k(x)|^2 
 \theta(\mu_d - \frac{k^2}{2m})
\end{equation}
and its equilibrium approximation 
$n_x^{\rm eq}(x) = \frac{1}{2}(n_x(x) + n_x(-x))$. Considering a parabolic 
potential barrier, these 1D densities can be expressed in dimensionless 
densities with $\xi=\sqrt{m\omega_x} x$. They are compared in 
Fig.\ref{densOEQ} for 
the cases $\mu_s-{\cal E}_0=0.4\omega_x,~\mu_d-{\cal E}_0=0$ and 
$\mu_s-{\cal E}_0=0,~\mu_d-{\cal E}_0=-0.4\omega_x$. The equilibrium 
approximation moves some electrons from the source ($x<0$) to 
the drain ($x>0$). Note that the equilibrium approximation
is exact at $x=0$.

Since $\psi_k^{(j)}(-x) = \psi_{-k}^{(j)}(x)$, the substitution
\eqref{subst} yields
\begin{equation}
	n_I(x,y) \to \sum_{\sigma = \pm 1} \sum_{j} 
\sum_{-\infty<k<\infty} |\psi_k^{(j)}(x)|^2|\chix^{(j)}(y)|^2 
\theta(\mu +\sigma \frac{V_{sd}}{2}- \frac{k^2}{2m})
\end{equation}
and similarly
\begin{equation}
n_{II}(x,y)	\to \frac{m}{2\pi}\sum_{\sigma=\pm 1} \Big(\mu + 
	\sigma \frac{V_{sd}}{2} - U(x,y) \Big)\theta \Big(\mu + 
	\sigma \frac{V_{sd}}{2} - U(x,y) \Big) \ .
\end{equation}
These formulae are the same as in the case of zero bias and
the source-drain splitting can be exactly mapped to a magnetic splitting by 
$eV_{sd} \leftrightarrow \g^*B$. Therefore, in the present Hartree
approximation, $\g_Q = \g^*$ exactly. This is consistent with the 
perturbative analysis presented in the main text.
\pagebreak
\section{Random Phase Approximation for the electron-electron interaction} \label{RPA}
\begin{figure}[ht]
{
 \includegraphics{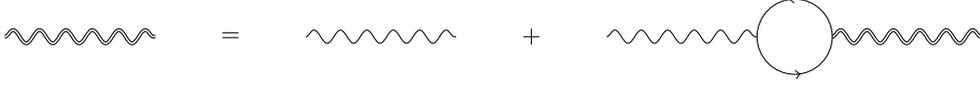}
}
\caption{
Diagrammatic representation of the Dyson self-consistent equation for 
the screened interaction in the RPA.
}
\label{dyson}
\end{figure}

We detail here a random phase approximation (RPA) calculation of the 
screened Coulomb interaction for electrons in the lowest transverse 
channel ($n=0$) and use it to justify the $\delta$-function model 
presented in the main text.

The 2D wavefunction for the lowest transverse channel is 
$\Psi_{k,0}(x,y) = \psi_{k,0}(x)\chi_0(y)$ with $\chi_0(y) = 
\left(\frac{m\omega_y}{\pi}\right)^{\frac{1}{4}}
e^{-\frac{m\omega_y y^2}{2}}$. The 1D bare Coulomb interaction is 
obtained by averaging over the transverse direction:
\begin{equation}
	V^c_0(x,x') = \frac{e^2}{4\pi\varepsilon_0\varepsilon_r}\int \ddd y 
	\int \ddd y' \frac{|\chi_{0}(y)|^2|\chi_{0}(y')|^2}
	{\sqrt{(x-x')^2 + (y-y')^2}} \ .
\end{equation}
In dimensionless lengths $\xi=x\sqrt{m\omega_x}$,
\begin{equation}
	V^c_0(\xi,\xi')= \omega_x \Lambda f(\xi-\xi',\frac{\omega_y}{\omega_x}) 
\end{equation}
with $\Lambda= \frac{e^2}{4\pi\varepsilon_0\varepsilon_r} 
\sqrt{\frac{m}{\omega_x}} \approx 3.35$ for typical experimental parameters 
in GaAs ($\epsilon_r = 13$, $m=0.07m_e$, $\omega_x \approx 1$meV) and
\begin{equation}
	f(\xi,\alpha) = \frac{\alpha}{\pi}\int \ddd \eta \int \ddd \eta' 
	\frac{e^{-\alpha (\eta^2 + \eta'^2)}}{\sqrt{\xi^2 + (\eta-\eta')^2}} 
	\qquad \to \qquad   \left\lbrace \begin{array}{ll}  \frac{{1}}{|\xi|} 
	& |\xi| \to \infty \\ & \\
 \sqrt{\frac{2\alpha}{\pi}}\left| \ln|\xi| \right| \qquad &  |\xi| \to 0
\end{array} \right. \ .
\end{equation}
Here $\alpha=\omega_y/\omega_x$.
Furthermore, because the QPC is not isolated, there will be an additional 
screening from the metal gates that will be important for large $|\xi|$. 
We model this qualitatively by using the bare interaction
\begin{equation}
	V^c_0(\xi,\xi')= \omega_x \Lambda f(\xi-\xi',\frac{\omega_y}{\omega_x})
	e^{-\frac{(\xi-\xi')^2}{\tau^2}} = \omega_x \tilde V^c_0(\xi,\xi')
\end{equation}
with $\tau=10$. The somewhat arbitrary choice of $\tau$ has, however, very 
little influence on the results presented below.

The screened interaction $V^c(\xi,\xi')=\omega_x \tilde V^c(\xi,\xi')$ 
is found by solving numerically the Dyson self-consistent equation
\begin{equation} 
	\tilde V^c(\xi,\xi')= \tilde V^c_0(\xi,\xi') +  \int \ddd \xi_1 \ddd 
	\xi_2 \tilde V^c_0(\xi,\xi_1) \Pi(\xi_1,\xi_2) \tilde V^c(\xi_2,\xi') \ .
\end{equation}
The RPA approximation of this relation is depicted diagrammatically in 
Fig.\ref{dyson}. 
In this approximation, the static polarisation reads, 
\begin{equation}
	\begin{split}
	 {\Pi}(\xi_1,\xi_2) &=
 \sum_{\sigma=\pm 1} 	\sum_{\delta_1,\delta_2=\pm }
 \int_{-\infty}^{\tilde\mu+\sigma \frac{\g^*B}{2\omega_x}} 
 \frac{\ddd \epsilon_1}{2\pi} 
 \int^\infty_{\tilde\mu+\sigma \frac{\g^*B}{2\omega_x}} 
 \frac{\ddd \epsilon_2}{2\pi} \frac{\phi_{\epsilon_2,\delta_2}(\xi_1) 
 \phi_{\epsilon_2,\delta_2}^*(\xi_2) \phi_{\epsilon_1,\delta_1}(\xi_2) 
 \phi_{\epsilon_1,\delta_1}^*(\xi_1)}{ \epsilon_1 -\epsilon_2}
	\end{split}
\end{equation}
where $\tilde \mu = \frac{\mu - U_{00}}{\omega_x}$ and the $\delta_{j}=\pm$ 
indicate the sign of $k$, i.e. the direction of the electron.

\begin{figure}[t]
{
 \includegraphics{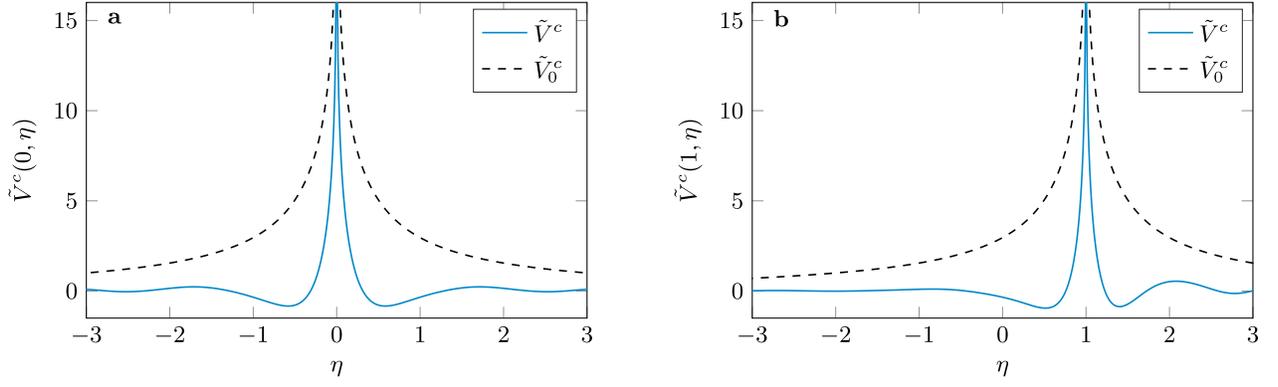}
}
\caption{
(a) Screened and bare dimensionless interactions between an electron at the 
centre of the QPC and an electron at position $\eta$ for $\tilde \mu = 0$, 
$B=0$ and $\omega_y=3\omega_x=3$meV. (b) Same as (a) but for an electron at 
$\xi=1$ and an electron at position $\eta$. 
}
\label{scrPot}
\end{figure}
Because the system is non-uniform, the polarisation is not uniform and 
$V^c(\xi,\xi')\neq V^c(\xi-\xi')$. The dimensionless interaction 
$\tilde V^c(0,\eta)$ between an electron at the centre of the QPC 
and an electron at position $\eta$ is compared to the bare interaction 
in Fig.\ref{scrPot}a for $\tilde \mu = 0$, $B=0$ and 
$\omega_y=3\omega_x=3$meV. 
The dimensionless interaction $\tilde V^c(1,\eta)$ between an electron at 
$\xi=1$ and an electron at position $\eta$ for the same parameters is 
shown in Fig.\ref{scrPot}b.

\begin{figure}[ht]
{
 \includegraphics{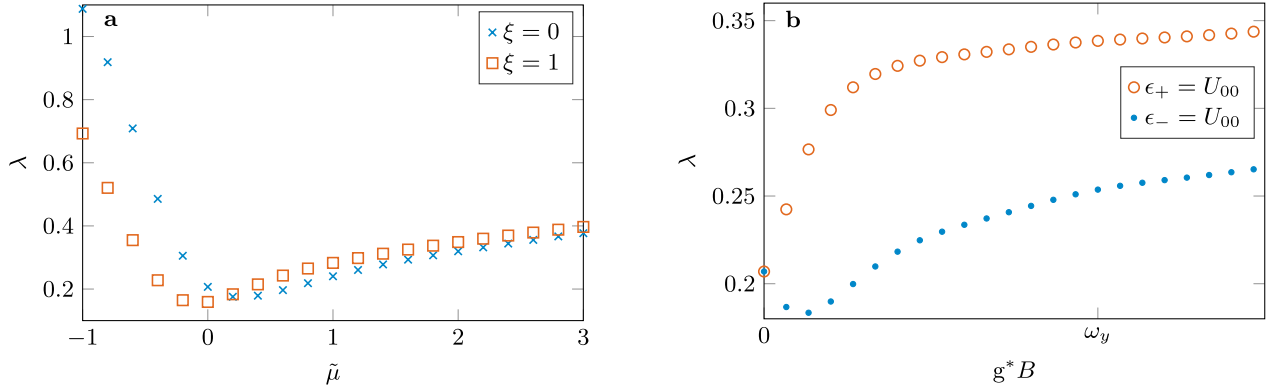}
}
\caption{
(a) Coupling constant $\lambda(\xi,\tilde \mu,B)$ at $\xi=0$ 
(centre of the QPC) and $\xi=1$ as a function of $\tilde \mu$ for 
 $\omega_y=3\omega_x=3$meV and $B=0$. (b) Coupling constant 
 $\lambda(\xi,\tilde\mu,B)$ 
at the centre of the QPC ($\xi=0$) as a function of the magnetic field 
$B$ at the positions of conductance steps. The condition $\epsilon_-=U_{00}$ 
yields the lower line going up (in e.g. Fig.3b of the main text) whereas the 
condition $\epsilon_+=U_{00}$ yields the lower line going down.
}
\label{lambda}
\end{figure}

The short-ranged screened interaction justifies 
the $\delta$-function approximation
\begin{equation}
V^c(\xi,\xi')\to \omega_x \pi^2 \lambda(\xi, \tilde \mu, B) \delta(\xi-\xi') 
\end{equation}
with the coupling constant
 \begin{equation}
 	\lambda(\xi,\tilde \mu, B) =\frac{1}{\pi^2}
 	\int_{-\infty}^\infty \tilde V^c(\xi,\eta)\ddd \eta \ .
 \end{equation}
The coupling constant at $\xi=0$ and $\xi=1$ is plotted as a 
function of $\tilde \mu$ in Fig.\ref{lambda}a for $B=0$.
 As $\tilde \mu$ becomes negative and the constriction depopulates, 
 the screening becomes less effective. For $\tilde \mu \lesssim -0.5$, 
 the contact approximation becomes questionable as the screened 
 interaction gets longer-ranged and eventually tends to the  bare 
 interaction. This is qualitatively different from the RPA result 
 for a uniform quantum wire which predicts an unphysical vanishing 
 coupling constant as the density vanishes (as a consequence of the 
 divergence of the density of state). This issue does not arise in 
 the present case of a QPC.

In the magnetic splitting, the step-positions (for the lowest transverse
channel) are obtained when $\epsilon_+=U_{00}$ 
(lower line going down in e.g. Fig.3b of the main text) or 
$\epsilon_-=U_{00}$ (lower line going up). The coupling constant $\lambda$ as 
a function of magnetic field $B$ with either one of these conditions 
satisfied is shown in Fig.\ref{lambda}b. This result applies 
also to the source-drain splitting ($\epsilon_\pm \to \mu_{s/d}$, 
$\g^*B \to V_{sd}$) by applying an analogous equilibrium approximation
as in section \ref{sec:ooe}. Remarkably, even though 
the coupling $\lambda$ can in principle vary significantly, as in 
Fig.\ref{lambda}a, it only varies by $\lesssim 20\%$ in the equations 
determining the position of the conduction steps in the source-drain 
bias spectroscopy experiment.  Furthermore, the interaction effect is 
most important for the lines going up and large $B$ (or $V_{sd}$), thus explaining 
why there is almost no perceptible difference in Fig.4c of the main text 
between the result with energy-dependent coupling (red dots) and with 
constant coupling $\lambda=0.25$ (black lines).

\end{document}